\documentclass{article}

\usepackage{PRIMEarxiv}

\usepackage[utf8]{inputenc} 
\usepackage[T1]{fontenc}    
\usepackage{hyperref}       
\usepackage{url}            
\usepackage{booktabs}       
\usepackage{amsfonts}       
\usepackage{nicefrac}       
\usepackage{microtype}      
\usepackage{lipsum}
\usepackage{fancyhdr}       
\usepackage{graphicx}       
\graphicspath{{media/}}     
\usepackage{xcolor}
\definecolor{myreferencecolor}{RGB}{0,0,255} 
\hypersetup{
    colorlinks=true,
    linkcolor=myreferencecolor,
    citecolor=myreferencecolor,
    urlcolor=myreferencecolor
}
\title{Student Debt and Behavioral Bias: A Trillion Dollar Problem
}

\author{
  Praful Raj \\
 The State University of New York at Buffalo \\
  \texttt{rajpraful166@gmail.com
} \\
}

\begin{document}
\maketitle

\begin{abstract}
 This literature review elucidates the implications of behavioral biases, particularly those stemming from overconfidence and framing, on the intertemporal choices made by students on their underline demand preferences for student loans. A secondary objective is to understand the potential utility of social media to assist students and young borrowers with the debt repayment process and management of their loan tenures. A close examination of the literature reveals a substantial influence of these behavioral and cognitive principles on the intertemporal choices made by students towards debt repayments. This affects not only the magnitude of loans they acquire but also the anchoring of the terms of loan conditions associated with repayment. Furthermore, I establish that harnessing social media as the potential to cultivate financial literacy and enhanced understanding of loan terms to expedite the process of debt redemption. This review could serve as a valuable repository for students, scholars, and policymakers alike, in order to expound on the cognitive biases that students and consumers often face when applying and entering into loan contract. 
\end{abstract}

\section{Introduction}
Student loans constitute the largest component of non-housing debt and the 2\textsuperscript{nd} largest liability after mortgages in the United States. As of Q4 2020, total outstanding student loans amounted to {\$}1.71 \cite{EducationDataInitiative} trillion held by over 42.9 million debtors (students), representing 7.31{\%} of the United States GDP \cite{FederalReserveBankofStLouis}. This exceeds the GDP of the major economies like Spain ({\$}1.40 trillion) and Australia ({\$}1.39 trillion) \cite{TradingEconomics}. Comparably, the {\$}1.9 trillion fiscal stimulus is predicted to increase GDP growth by 2{\%} points and spur inflation through heightened aggregate demand \cite{CRS}. Over one million student loans enter default annually, with averages of 15{\%} in default and 11{\%} of graduates defaulting in their first year \cite{EducationDataInitiative}. Student loan debt is dramatically growing 6 times faster than the nation’s economy \cite{EducationDataInitiative}. Elevated student debt burdens are associated with reduced community well-being, lower home ownership rates, decreased entrepreneurship, delayed family formation, reduced retirement savings, higher bankruptcy risks, and declining net worth \cite{ElliottLewis2015, KrishnanWang2018}. \\

Classical consumer theory assumes individuals are rational utility maximizers exhibiting consistent, optimal decision-making \cite{Debreu1959}. However, empirical evidence reveals consumers frequently deviate from perfect rationality due to bounded rationality and cognitive biases \cite{Simon1955, Kahneman2003}. Real-world choices often reflect heuristics, framing effects, and context-dependent preferences inconsistent with wealth maximization \cite{ThalerR1980, Tversky1981}. These anomalies violate classical assumptions, indicating consumers do not uniformly conform to homo economicus predictions \cite{Camerer1998}. Academic literature has elucidated various cognitive biases which distort ostensibly rational decision-making, including choice architecture, confirmation bias, overconfidence, framing effects, mental accounting, peak-end bias, and decision fatigue \cite{ThalerR2008, Nickerson1998,MooreDA2008, Tversky1981, Thaler1999, Kahneman2000, Vohsbaumeister2008}. This paper will focus on overconfidence and framing effects on the decision-making processes and revealed preferences of student borrowers, while controlling for demographic factors. 
Overconfidence is a belief of performing better than one’s preparedness or understanding in a particular situation. Research has shown that overconfidence has led to over-optimism, unrealistic choices, and wrong decisions leading to unexpected results. When some kind of gain is involved between the two same choices but presented in a different way, resulted participants to choose conservative option and that involves less risk. But when same task is presented involving some kind of loss, participants tend to select the riskier choice \cite{tversky1986rational,tversky1991loss,Kahneman2003,kahneman2007frames}. Research has also shown that there is an aversion to debt due to labelling (framing) contract as a “loan” \cite{Caetanoy2019}. These biases affect students' borrowing decisions, preventing them from making rational choices. Given the proliferation of social media and its ability to shape decisions through virtual connections, these platforms may represent an effective mechanism to inform student borrowing choices. This review will analyze how social media could potentially impact future debt accrual, living standards, and other dimensions of the student loan crisis. By publicizing the pros, cons, and externalities of student loans through viral information diffusion, social networks may ameliorate suboptimal borrowing behavior grounded in systemic biases. 

\section{Overconfidence}
The Titanic tragedy exemplifies overconfidence biases, as the captain's hubris led him to ignore iceberg risks. Similarly, overconfidence among US policymakers enabled initial coronavirus spread by downplaying pandemic threats. Additionally, overconfidence among Chernobyl scientists regarding safety precautions catalyzed the disaster. These examples illustrate that overconfidence is an enduring cognitive bias which distorts risk evaluation and impedes optimal outcomes. Bounded rationality limits the ability of individuals to objectively calculate expected utility, leading to suboptimal decision-making \cite{Simon1955, Williamson1981}. Consequently, principal-agent problems arise wherein asymmetric information among actors creates a moral hazard that hinders welfare maximization \cite{Eisenhardt1989,Jensen1976}.\\
Overconfidence, defined as “overestimation of one’s actual performance” or “excessive precision in one’s belief” \cite{MooreDA2008} has been studied by economists at least as early as Adam Smith \cite{CitationExample2}, who wrote in The Wealth of Nations: “the overweening conceit which the greater part of men has their own abilities, is an ancient evil remarked by the philosophers and moralists of all ages”. The paper “Paying not to go to the gym”, discusses the question “how do consumers choose from a menu of contract?" DellaVigna {\&} Malmedier \cite{CitationExample3}, cites “Overconfidence about future self-control or future efficiency. They found that we “overestimate the attendance as well as the cancellation probability of automatically renewed contract”. We end up paying more compared to the charge based on per visit \cite{CitationExample3}. In fact, in taking health club membership, we do not bother to go even after paying a significant amount \cite{CitationExample3}. This irrational behavior impacts our crucial future decisions like borrowing, purchasing houses, cars, savings, and insurance, among others. Students may be the victims of overconfidence and fail to see their prospects leading to an inability to manage their finances well. Similarly, students are also overconfident about their future self-control and future efficiency in taking borrowing decision.
\subsection{Overconfidence in Financial Behavior}
According to the National Financial Capability Study (NFCS), an in-depth examination of the financial behaviors and knowledge of the millennial cohort reveals certain alarming patterns. \cite{CitationExample7} utilized this data to analyze the financial tendencies prevalent within this demographic. A predominant observation was the significant indebtedness of millennials, which was primarily attributed to credit card balances, educational loans, and expenditures stemming from upscale lifestyles. This study underscores a critical concern: millennials often exhibit an inflated perception of their own financial competency. This overestimation can potentially transition to financial overconfidence, an attitude which could be detrimental when making pivotal economic decisions. Pertinently, such overconfidence has been correlated with an escalation in student loan commitments and a consequential uptick in delinquency occurrences. \cite{CitationExample5} reports that more than half of the millennials burdened with student loans harbor apprehensions regarding their repayment capabilities. Furthermore, an alarming 60{\%} of millennials have been identified to adopt costly credit card management practices. An additional area of concern pertains to the alternate financial services that are gaining traction among millennials. Specifically, \cite{CitationExample5} observes that approximately 43{\%} of millennials have availed of at least one form of such services, which includes pawnshops, payday lenders, and rent-to-own establishments. It is imperative to note that the escalation in student loan assumption post-secondary education is not solely an outcome of deficient financial literacy but can also be attributed to an unwarranted confidence in one's financial acumen concerning loans and related expenditures.

\subsection{Overconfidence borrowing behavior}
 Optimal student borrowing necessitates comprehending the dynamics of interest rates on debt servicing and financial management. While federal student loan interest appears as low as 4.50{\%}\footnote{Undergraduate: 2.75{\%} p.a. (Direct subsidized loans {\&} Direct unsubsidized loans). Graduate or Professional: 4:30{\%} p.a. (Direct unsubsized loans). Parent and Graduate or Professional students: 5:30{\%} p.a. (Direct PLUS Loans). Perkins loans (regardless of first disbursement date) have a fixed interest rate of 5{\%}. Effective on or after July 1 2020, and before July 1, 2021. All interest rate are fixed rates that will not change for the life of the loan.} annually, compounding effects over time are considerable. In the paper, “young adults’ financial literacy and overconfidence bias in debt markets”\cite{CitationExample6}, a negative relationship between financial behavior and overconfidence emerges. “Overconfident respondents were more likely to borrow as compared to others”. In a sample study of 600 adults aged between 18 and 35, the quantitative measure of overconfidence was found to be significantly and positively related to unhealthy debt behavior \cite{CitationExample6}. In the study of mortgage payment delinquency using the 2015 National Financial Capability Study data set, it was observed that overconfident people have a likelihood of delinquency three times as high as those with a higher objective and subjective literacy \cite{CitationExample7}. Empirical analysis of UK households demonstrates a negative correlation between financial literacy and consumer credit reliance, with less financially literate consumers exhibiting greater participation in credit markets \cite{Lusardi2015}.  They have a higher level of payday loans and home collected credit \cite{CitationExample8}. It is essential in terms of policy implications and means to formulate more effective financial programs \cite{CitationExample9}.

\subsection{Income expectations}
\cite{CitationExample10}, examined a particular behavioral bias i.e., overconfident income expectations, and analyzed how it influences borrowing behavior. It was found that participants with higher income expectations consume more, take up more debt, and keep the higher level of debt throughout the experiment. In particular, overconfident participants run the risk to stay in debt. This behavior can come in contact with other behavioral biases such as time inconsistency, hyperbolic discounting, and self-control while taking the decision of getting in the line of federal student loans. Students overestimate their ability to earn more and CEOs overestimate their ability to run the company in a better way, both in order to repay debts on time. Overconfidence with higher expectations is also seen in entrepreneurial behavior. As per one survey conducted in 18 countries, a significant negative correlation between the reported level of entrepreneurial confidence and the approximate survival chance of nascent entrepreneurs across countries has been found \cite{CitationExample11}. In one of the experimental entry games, “individuals tend to overestimate their chances of relative success and enter more frequently”, players payoff was based on their skills. Surprisingly, when subjects were asked to enter into the experimental sessions on their own, they were even more overconfident \cite{CitationExample12}.
\subsection{Overconfident CEOs}
Overconfidence among corporate executives can engender suboptimal financial and operational decisions. For instance, overconfident CEOs exhibit a penchant for excessive leverage via short-term debt financing, which can jeopardize company performance \cite{Malmendier2005}. They tend to also get fired by their board of directors \cite{Goel2008}. One of the reasons being that overconfident CEOs pledge more collateral in order to lower the loan rate. In the paper, “confidence, and overconfidence in banking”\cite{CitationExample13}, it was found that “financial crisis was determined mainly by the increased confidence of the mid-confident bank CEOs and not the behavioral biases of overconfident CEOs”. A large sample of data was taken from American Bank in the period 2000-2013 to study one of the reasons for the financial crash of 2007- 2008. The results showed a negative correlation between overconfident behavior and the bank's value\cite{CitationExample13}.\\
Many studies have shown that a CEO's overconfidence has significant consequences on the company's investment and financing decisions, company performance, future growth, and development. A study by \cite{CitationExample14}, showed that overconfident CEOs tend to increase the degree of leverage and increase the number of loans, especially short-term loans. It is a high chance that more leveraged and aggressive companies have overconfident CEOs. If students are similarly overconfident as CEOs are, we can expect them to have high levels of debt and to have a more difficult time repaying it. Over a million students loans enter default each year. An average of 15{\%} of student loans are in default at any given time. 11{\%} of student loan holder default within their first year of graduation. \\
\cite{CitationExample15} and \cite{CitationExample16}, illustrated that people are more optimistic about their future life events. In the study of a competence driver as compared to group of drivers, most of the drivers claimed themselves as a skillful and less risky compared to their pool of rest of the drivers \cite{CitationExample17}. Similarly, students also tend to be optimistic about positive future experiences and pessimistic about negative events that they would have in comparison to their peers.\\
This section hypothesizes overconfidence substantially influences student borrowing decisions, consistent with evidence across domains that this bias distorts choice. As a prevalent anomaly challenging rational decision-making, overconfidence merits investigation regarding its impact on suboptimal student loan accrual.

\section{Framing}
The modern information economy provides a profusion of data in myriad formats, conferring both benefits and costs for decision-makers. While abundant information should theoretically improve choices, bounded rationality implies cognitive limits may hinder processing and comprehension. When our decision gets influenced by the way information is presented it is called a “Framing Effect”. It has been demonstrated empirically that framing of the desired perceived information in a certain way has a direct impact on decision making \cite{tversky1981framing, mullainathan2008psychology, banks1995understanding}.\\
\cite{CitationExample18} demonstrated in their paper that framing can have implications on the marginal propensity to consume and spend. An increase in income that is framed as a gain (or rebate) from the current income level is more likely to be spent as compared to when an increase in income is framed as a return to a previous state of income, which is more likely to be saved.\\

For federal student loans, students face cognitive constraints in processing the extensive terms and conditions, implying full comprehension is improbable. The detailed contractual fine print exceeds bounded rationality limits, resulting in confusion despite disclosure \cite{CitationExample19}. Empirical evidence indicates suboptimal comprehension of student loan terms among undergraduate borrowers. Data linking survey responses to administrative records demonstrate many students neither accurately estimate college costs nor understand accumulated debt burdens, despite loan disclosures \cite{Boatman2017}.\\

Income-driven repayment (IDR) plans represent a critical contractual clause, lowering default risk by linking payment obligations to earnings contingencies. IDR protects borrowers facing repayment difficulties due to unemployment or persistently low wages by providing payment flexibility. The repayment amount is attached with the level of income and keeps changing accordingly.\cite{CitationExample20}, studies framing bias with respect to IDR by surveying University of Maryland undergraduates. Some undergraduates were presented with the option of IDR plans that were not framed (neutral) while others were randomly assigned the plans with emphasized framing on the word IDR cost. It was found that those with the neutral plans were 50{\%} less likely to choose IDR plan, while those who were randomly assigned the plans with emphasized framing on IDR insurance were 60{\%} more likely to choose the IDR. Emphasizing the word “insurance” while framing the information was thus found to have a significant impact on the decision.\\

\cite{CitationExample21}, through a randomized control trial of over six thousand high school seniors, community college students, and adults without a college degree found both the framing and labeling effect on students' preferences who have plans for taking loans for future studies. Simply replacing the word with "loan" reduces the likelihood of selecting those options by 8-11{\%} among the samples. Talking particularly about the community of Black and Hispanic, the results were more pronounced. They were twice as likely to avoid loans than white students.\\

In the context of financial aid, the Direct PLUS Loan stands as a pivotal component for graduate and professional students. It represents a federal student loan that is directly applied to the recipient's school account, covering essential expenses such as tuition, fees, and room and board for on-campus students. For instance, in the application form for Direct PLUS Loan for Graduate/Professional Students, applicants are presented with options that, when selected, authorize the school to use the Direct PLUS Loan to settle other education-related charges incurred at the institution. These may encompass a variety of expenses, from library fines and campus parking tickets to lab fees and charges for minor damages to school property. This authorization, while not mandatory, provides a convenient means for students to manage their financial obligations to the institution without the need for separate transactions or additional loans \cite{FederalStudentAid}. 
This loan is acceptable, but the option given for the loan amount requested is framed to encourage students to go for a higher (maximum) amount.\\
\textbf{Option 1}: I want to borrow the maximum Direct PLUS Loan 
amount for which I am eligible, as determined by the schools.\\
\textbf{Option 2}: I would like to specify a loan amount.\\ 
\textbf{Option 3}: I don't know the amount I want to borrow. I will contact the school.\\

    Option 1 is inadvisable because it may discourage financial self-reliance. Even students considering bootstrapping or part-time work to cover expenses could opt for unnecessary loans instead. The framing may implicitly encourage irresponsible behaviors like late book returns or property damage by those previously unwilling to incur such costs. This could incentivize unjustified borrowing, contradicting the aim of minimizing student debt accrual.\\

    \cite{Caetanoy2019} demonstrated that the mere labeling of a contract as a “loan” can induce aversion to debt, decreasing its likelihood of being selected over a financially equivalent contract by over 8{\%}. This finding was further corroborated by a randomized control trial conducted by \cite{CitationExample22}, which revealed that providing high school seniors with information about federal loans decreased their propensity to borrow. However, the same study found that disseminating information about income-based repayment (IDR) options positively influenced borrowing decisions. This is a crucial insight as it suggests a viable strategy for improving college access among low- and middle-income populations, who typically exhibit aversion to borrowing. By reframing the narrative around student loans and emphasizing the availability and benefits of IDR options, it may be possible to mitigate the fear associated with borrowing and, consequently, broaden access to higher education for these demographics.\\

    Framing information can have both negative and positive impacts - it depends on the intention of the information provider behind the communication and the way he/ she wants the information to be perceived by the information seeker. Salient information has a positive impact on student behavior as can be seen in another study done by \cite{CitationExample23}. Empirical evidence shows students receiving communications about their borrowing patterns exhibit improved academic performance, including higher course loads and GPAs in subsequent terms \cite{Darolia2018}.\\

    Applying the principles to the domain of retirement savings, a paper by \cite{Benartzi2011} shows that the design of the retirement savings vehicle has a vital impact on savings rate and investment decisions. “Lifecycle funds do reshape individual portfolio choices through large default and framing effect \cite{CitationExample24}”. In 401(K) participation decision, prior to automatic enrollment, the default is non-participation, while under the automatic enrollment, the default is participation. There was a change of information in automatic enrollment so “The “default” behavior appears to result from participant inertia and from employee perceptions of the default as investment advice” \cite{CitationExample25}. \\
    
    Thus, it can be inferred that inadequate financial literacy and framing can hamper the students from taking a rational decision with respect to their borrowing needs. What is also noteworthy about framing is that it is as true for amateurs, much like we discussed, as it is for financial experts. In the case of auditors, they failed to detect the error in financial statements when presented with the information that was deliberately framed for the framing experiment \cite{CitationExample26}.\\

\section{Social Media}
Belongingness represents a fundamental human motivation \cite{Baumeister1995}. Social media platforms satisfy this need for affiliation and connection \cite{Seidam2013}. Though, Social media confers praiseworthy utility gains that continue to expand user adoption.
 An experiment by famous social psychologist, \cite{CitationExample27} shed light on the universal human behavior, i.e., conformity. When people were asked to make a decision on their own, without seeing what others were doing, people made the right choices and gave the correct answers. But when everyone else gave an incorrect answer, people made a mistake in more than one-third instances. “Conformity experiments have been replicated and extended in more than 130 experiments from seventeen countries,including Zaire, Germany, France, Japan, Norway, Lebanon, and Kuwait” \cite{Adkisson2008}. Social media can help to solve various problems without forcing people to do so, through Libertarian Paternalism, “an approach that preserves freedom of choice but that authorizes institutions to steer people in directions that will promote their welfare” \cite{Thaler2003}.\\

The impact of social media and television advertisement can be illustrated through this example. In 1985, the Texas Department of transportation was looking for means to reduce the litter on the roadways. GSD \& M Idea City LLC \cite{GSDM} came up with the slogan “Don’t mess with Texas”. For television ads, state enlisted popular Dallas Cowboys football players participated, especially targeting the unresponsive audience. Within the first year of the campaign, litter in the state was reduced by a remarkable 29 \%, and in its first six years, there was a 72 \%  reduction in visible roadside litter.\\

We can apply the concept of herd behavior that comes from different sources such as friends, families among other. Many times, our investment decisions are being influenced by our friends, families and colleague’s portfolio. In one of the experiments in San Marcos, information on energy consumption was given to 300 families of California along with their neighborhood of same time period i.e., past few weeks. After few weeks, it was observed that there was significant decrease in the energy consumption of a family whose consumption was above average in last few weeks and vice versa in other case. This is a boomerang effect \cite{Nudge2008}.\\

Social networks exert normative influence on health behaviors, as empirical research indicates young adults perceive their health attitudes and actions are shaped by social media content \cite{Vaterlaus2015} Through the Qualitative analysis, young adult’s health behavior (diet {\&} exercise) got influenced using social ecological framework. Participants agreed that technology can be both a motivation and barrier by posting statuses and pictures online. The study of student behavior revealed that social networking services can enhance bonding and can encourage engagement to studies and create learning environment and peer support \cite{CitationExample31}. Students borrowing facts must be publicized like the stock market to induce positive sentiments in order to push the economy. Social media platforms can positively motivate students to pay their debts on time.\\
The boomerang effect could potentially serve as a tool to incentivize students to enhance their borrowing and loan repayment conduct. Federal dissemination of student loan statistics and repayment norms via social media could incentivize timely debt servicing. Communicating facts about collection rates, disbursement, defaults, and outstanding balances frames student borrowing as a civic duty, potentially decreasing delinquency. As an illustration, the dissemination of the fact that these loans constitute taxpayer funds earmarked for the bright future of students and the nation, thereby invoking a sense of responsibility towards fellow taxpayers, may stimulate students to settle their outstanding educational loans. This, in turn, would likely foster an increased propensity for savings while concurrently reducing the propensity for expenditure.\\

\section{Conclusion} 
This paper conducts a comprehensive literature review encompassing various scholarly works exploring the ramifications of overconfidence and framing on diverse facets of decision-making. Particular emphasis is placed on the deployment of social media platforms as a mechanism for
ameliorating biases at a broad scale. The paper scrutinizes the intricacies of overconfidence behavior across multiple dimensions, including financial and lending conduct, as well as
overconfidence within the realm of corporate executive leadership. The consistent empirical observation posits that overconfidence exerts a deleterious influence on decision-making, with no empirical substantiation for the contention that the advantages conferred by overconfidence behavior outweigh the attendant detriments or losses. Furthermore, the examination of framing extends to various contexts encompassing perceptions of gains and losses, portfolio allocation, debt and credit decisions, and participation in 401(K) retirement plans. Grounded in the empirical evidence, we posit that framing may facilitate more astute decision-making among students. In the paper's concluding segment, we proffer social media as a tool for mitigating cognitive biases. This proposition is fortified by instances where policymakers and governmental entities have effectively harnessed these platforms to address societal issues without imposing decisions upon the populace. In summation, this paper underscores the imperative of addressing the pervasive impact of overconfidence and framing on decision-making processes, whilst proposing innovative remedies to debias these cognitive
inclinations.\\

\bibliographystyle{unsrt}  
\bibliography{references}
\end{document}